\documentclass[aps,11pt,nofootinbib,preprintnumbers,groupedaddress,superscriptaddress]{revtex4-2}

\usepackage{amssymb, amsfonts, amsmath, bm}
\usepackage[
]{graphicx}
\usepackage{color}
\usepackage{empheq}

\usepackage[
]{hyperref}
\hypersetup{%
 setpagesize=false,
 bookmarksnumbered=true,%
 bookmarksopen=true,%
 colorlinks=true,%
 linkcolor=blue,%
 citecolor=red}

\def\ii{i\hspace{-.1em}i}
\def\iii{i\hspace{-.1em}i\hspace{-.1em}i}
\def\iv{i\hspace{-.1em}v}

\begin{document}
\title{Photon emission from inside the innermost stable circular orbit}
\author{Takahisa Igata}
\email{igata@post.kek.jp}
\affiliation{KEK Theory Center, 
Institute of Particle and Nuclear Studies, 
High Energy Accelerator Research Organization, Tsukuba 305-0801, Japan}
\author{Kazunori Kohri}
\email{kohri@post.kek.jp}
\affiliation{KEK Theory Center, 
Institute of Particle and Nuclear Studies, 
High Energy Accelerator Research Organization, Tsukuba 305-0801, Japan}
\affiliation{The Graduate University of Advanced Studies (SOKENDAI), Tsukuba 305-0801, Japan}
\affiliation{Kavli Institute for the Physics and Mathematics of the
  Universe (WPI), University of Tokyo, Kashiwa 277-8583, Japan}
\author{Kota Ogasawara}
\email{kota@tap.scphys.kyoto-u.ac.jp}
\affiliation{Theoretical Astrophysics Group, Department of Physics, Kyoto University, Kyoto 606-8502, Japan}
\date{\today}
\preprint{KEK-Cosmo-0273}
\preprint{KEK-TH-2305}
\preprint{KUNS-2858}

\begin{abstract}
We consider a situation where a light source orbiting 
the innermost stable circular orbit (ISCO) of the Kerr black hole is 
gently falling from the marginally stable orbit due to an infinitesimal perturbation.
Assuming that the light source emits photons isotropically, 
we show that the last radius at which more than 50\% of emitted photons 
can escape to infinity is approximately halfway 
between the ISCO radius and the event horizon radius.
To evaluate them, we determine emitter orbits 
from the vicinity of the ISCO, 
which are uniquely specified for each black hole spin, 
and identify the conditions for a photon to escape from any point 
on the equatorial plane of the Kerr spacetime to infinity 
by specifying regions in the two-dimensional photon impact parameter space completely. 
We further show that the proper motion of the emitter 
affects the photon escape probability and blueshifts the energy of emitted photons. 
\end{abstract}
\maketitle

\section{Introduction}
\label{sec:1}
In recent years, significant progress has been made 
in observing a massive compact object located at the center of a galaxy.
The shadow observation of the M87 galactic center 
suggests that the central object is a black hole~\cite{Akiyama:2019cqa}.
In principle, we cannot receive any signals coming from inside the black hole 
because nothing can escape from it, 
but we can observe phenomena just outside the event horizon.
In other words, all the observations of black holes, including the shadow observations, 
without exception focus on unique phenomena in the vicinity of the horizon.
However, naively, the strong gravitational sphere of a black hole does not allow 
most signals emitted from the region close to the horizon to escape. 
Therefore, some mechanisms to increase the escape probability of such signals are 
necessary to observe unique phenomena near the horizon~\cite{Igata:2019pgb}.

Recently, the escape probability of photons emitted isotropically 
from an emitter orbiting a circular geodesic near a Kerr black hole 
was shown to be more than 50\% for an arbitrary black hole spin parameter 
and an arbitrary orbital radius~\cite{Igata:2019hkz,Gates:2020els}.
Furthermore, depending on the direction of the emission, 
the Doppler shift cancels the gravitational redshift, and as a result, 
the signals can reach 
a distant observer in observable bands~\cite{Cunningham:1973,Igata:2019hkz,Gates:2020sdh}.
The mechanism common to both phenomena is a relativistic boost or 
beaming due to the proper motion of the emitter.
In recent years, such the relativistic effects have been actively investigated 
through observations of a star orbiting around the center of 
our galaxy~\cite{Abuter:2018drb,Saida:2019mcz,Takamori:2020ntj}.
Even though the radius of the innermost stable circular orbit (ISCO) 
arbitrarily approaches the horizon radius in the maximum spin limit, 
the escape probability still exceeds a half~\cite{Igata:2019hkz,Gates:2020els}, 
and the Doppler blueshift is still dominant at least for the forward emission, 
suggesting that the emitter on the ISCO is a useful probe of the near-horizon.
In fact, a light source that seems to be orbiting 
the ISCO was observed~\cite{Abuter:2018}.
In particular, in the Atacama Large Millimeter/submillimeter Array observations of 
Sagittarius A*—a black hole candidate with an accretion disk 
at the center of our galaxy—a short-timescale variation 
of the radio flux density was reported~\cite{Iwata:2020pka}. 
This variation timescale ($\sim 30~\mathrm{minutes}$) is comparable 
to the orbital period of the Schwarzschild ISCO ($\sim 3 r_{\mathrm{g}}$, 
where $r_{\mathrm{g}}$ is the Schwarzschild radius). 
We can interpret this phenomenon as a sign that a hot spot is 
orbiting on the inner edge of the accretion disk 
and as the appearance of the relativistic beaming effect due to orbiting at high speed.

If an object orbiting the ISCO, such as a hot spot, is perturbed slightly, 
it must transit from the marginal orbit to a plunge orbit into the center.
If we continue to receive signals from the falling emitter, 
to what depth of the gravitational potential will we achieve the observation?
In other words, if we assume that the escape probability of a photon is 
50\% or more as an indicator of observability, 
what is the last radius where the probability becomes a half?
This threshold value is closely related to 
the unstable photon circular orbit~\cite{Synge:1966okc} 
and appears in the context of the loosely trapped surfaces~\cite{Shiromizu:2017ego}, 
which are the recent development of photon 
surfaces~\cite{Claudel:2000yi,Galtsov:2019bty,Koga:2020akc}.
The purpose of this paper is to identify some thresholds related to 
the photon escape probability if the emitter \textit{gently} leaves 
the ISCO and falls into the black hole.

This paper is organized as follows. 
In Sec.~\ref{sec:2}, 
we review a formulation of particle motion in the Kerr spacetime. 
In Sec.~\ref{sec:3}, clarifying necessary and 
sufficient escape conditions for photons 
emitted from the equatorial plane, 
we completely specify the photon escape parameter region 
in the space of impact parameters.
In Sec.~\ref{sec:4}, we describe the dynamics of an emitter 
plunging from the ISCO into the horizon. 
In Sec.~\ref{sec:5}, formulating the escape probability of a photon 
measured in the rest frame of the emitter, 
we evaluate the photon escape probability for each black hole spin and 
discuss some properties relevant to the emitter dynamics. 
Section~\ref{sec:6} is devoted to a summary and discussions. 
In this paper, we use units in which $G=1$ and $c=1$.

\section{Geodesic system in the Kerr spacetime}
\label{sec:2}
The Kerr spacetime metric in the Boyer-Lindquist coordinates is given by
\begin{align}
\label{eq:metric}
&g_{\mu\nu}\:\!\mathrm{d}x^\mu\:\!\mathrm{d}x^\nu
=-\frac{\Delta \Sigma}{A}\:\!\mathrm{d}t^2
+\frac{\Sigma}{\Delta}\:\!\mathrm{d}r^2
+\Sigma\:\!\mathrm{d}\theta^2
+\frac{A}{\Sigma}\sin^2\theta \left(
\mathrm{d}\varphi-\frac{2Mar}{A}\:\!\mathrm{d}t
\right)^2,
\\
&\Delta=r^2-2Mr+a^2,
\quad
\Sigma=r^2+a^2\cos^2\theta,
\quad
A=(r^2+a^2)^2-a^2 \Delta \sin^2\theta,
\label{eq:metfcts}
\end{align}
where $M$ and $a$ are mass and spin parameters, respectively. 
We assume that the metric denotes a black hole spacetime, i.e., 
$0\leq a\leq M$. Then, the event horizon is located at 
$r=r_{\mathrm{H}}:=M+\sqrt{M^2-a^2}$.
The metric admits a stationary Killing vector $\partial/\partial t$,
an axial Killing vector $\partial/ \partial \varphi$,
and a nontrivial rank-2 Killing tensor~\cite{Walker:1970un}
\begin{align}
\label{eq:KT}
K_{ab}=\Sigma^2 (\mathrm{d}\theta)_a(\mathrm{d} \theta)_b
+\sin^2\theta
\left[\:\!
(r^2+a^2) (\mathrm{d}\varphi)_a-a\:\!(\mathrm{d}t)_a
\:\!\right]
\left[\:\!
(r^2+a^2) (\mathrm{d}\varphi)_b-a\:\!(\mathrm{d}t)_b
\:\!\right]-a^2 \cos^2\theta g_{ab}.
\end{align}

We consider a freely falling particle system (i.e., a geodesic system) 
in the Kerr spacetime. 
Let $p_a$ be a canonical momentum of a particle, 
and let $H$ be the Hamiltonian of affinely parametrized geodesic system, 
\begin{align}
H=\frac{1}{2} g^{ab}p_a p_b=\frac{1}{2}\left[\:\!
\frac{\Delta}{\Sigma} \:\!p_r^2
+\frac{p_\theta^2}{\Sigma}
+\frac{\Sigma}{A\sin^2\theta} \:\!p_\varphi^2
-\frac{A}{\Delta\Sigma} \left(
p_t+\frac{2Ma r}{A} \:\!p_\varphi
\right)^2
\:\!\right],
\end{align}
where $g^{ab}$ is the inverse metric. 
Since $t$ and $\varphi$ are cyclic variables in $H$, 
the conjugate momentum components $p_t=-E$ and $p_\varphi=L$ 
are constants of motion. We can interpret $E$ and $L$
as conserved energy and angular momentum along a geodesic, respectively. 
From reparameterization invariance of an affine parameter, 
a particle must satisfy the constraint condition
\begin{align}
\kappa=-g^{ab}p_a p_b,
\end{align}
where $\kappa=1$ for a massive particle with unit mass, 
and $\kappa=0$ for a massless particle. 
From the existence of the Killing tensor~\eqref{eq:KT}, 
a particle has another quadratic constant of motion in $p_a$, 
the so-called Carter constant, 
\begin{align}
Q=K^{ab}p_a p_b-(L-a E)^2.
\end{align}
The four constants of motion commute with each other in the Poisson brackets. 
Therefore, the equation of motion (i.e., the geodesic equation) 
in the Kerr spacetime is integrable in the Liouville sense.

Not only is it integrable, but it causes the separation of variables 
in the Boyer-Lindquist coordinates 
by using the Hamilton-Jacobi method~\cite{Carter:1968rr}.
The separated equations in the canonical variables are rewritten
by the derivatives $\dot{x}^\mu$ with respect to the affine parameter as
\begin{align}
\label{eq:tdot}
\Sigma \:\!\dot{t}&=a (L-aE\sin^2\theta)+\frac{r^2+a^2}{\Delta} \left[\:\!
(r^2+a^2)E-a L
\:\!\right],
\\
\label{eq:phidot}
\Sigma \:\!\dot{\varphi} &=\frac{L-aE \sin^2\theta}{\sin^2\theta}+\frac{a}{\Delta}\left[\:\!
(r^2+a^2) E-a L
\:\!\right],
\\
\label{eq:rdot}
\Sigma\:\! \dot{r}&=\sigma_r \sqrt{R_\kappa},
\\
\label{eq:thetadot}
\Sigma \:\!\dot{\theta}&=\sigma_\theta \sqrt{\Theta_\kappa},
\end{align}
where $\sigma_r=\mathrm{sgn}(\dot{r})$, 
$\sigma_\theta=\mathrm{sgn}(\dot{\theta})$, and
\begin{align}
R_\kappa(r)&=\left[\:\!
(r^2+a^2) E- a L
\:\!\right]^2-\Delta \left[\:\!
(L-a E)^2+Q+\kappa \:\!r^2
\:\!\right],
\\
\label{eq:THETA}
\Theta_\kappa(\theta) &=Q-\cos^2\theta \left[\:\!
\frac{L^2}{\sin^2\theta}+a^2(\kappa -E^2)
\:\!\right].
\end{align}
Note that $R_\kappa$ and $\Theta_\kappa$ must be nonnegative for physical solutions, 
as is seen from Eqs.~\eqref{eq:rdot} and \eqref{eq:thetadot}. 
We use units in which $M=1$ in what follows.

\section{Photon escape conditions}
\label{sec:3}
We consider 
escape conditions of a photon (i.e., $\kappa=0$) emitted from a point on 
the equatorial plane.
Let us denote a momentum of a photon as $k_a$ instead of $p_a$. 
Since only considering a photon escaping to infinity, 
without loss of generality, we may assume that $E>0$. 
It is useful to introduce impact parameters 
\begin{align}
b=\frac{L}{E}, \quad q=\frac{Q}{E^2}.
\end{align}
Rescaling $k^a$ as $k^a/E\to k^a$, 
we obtain a two-parameter family of the null geodesic equations 
parametrized by $(b, q)$.

Let $\theta_*$ be the polar angle coordinate value 
of the emission point, i.e., $\theta_*=\pi/2$.
The function $\Theta_0$ in Eq.~\eqref{eq:THETA} evaluated there leads to 
\begin{align}
\Theta_0(\theta_*)
=q-\cot^2\theta_* (
b^2-a^2\sin^2\theta_*)
=q \geq 0.
\end{align}
Hence, a photon emitted from the equatorial plane must always 
have a nonnegative value of $q$. 
Hereafter, we assume that $q\geq 0$.

Let us clarify the escape conditions of a photon, 
which are determined from the radial equation of motion~\eqref{eq:rdot}.
The function $R_0$ is factored as 
\begin{align}
R_0&=\left[\:\!
r^2+a(a-b)
\:\!\right]^2-\Delta \left[\:\!
q+(b-a)^2
\:\!\right]
\\
&=r(2-r) (b-b_1)(b-b_2),
\end{align}
where 
\begin{align}
\label{eq:b1}
b_1(r;q)&=\frac{-2a r+\sqrt{r\Delta \left[\:\!
r^3-q(r-2)
\:\!\right]}}{r(r-2)},
\\
\label{eq:b2}
b_2(r;q)&=\frac{-2a r-\sqrt{r\Delta\left[\:\!
r^3-q(r-2)
\:\!\right]}}{r(r-2)},
\end{align}
which show the values of $b$ at a radial turning point. 
Note that $b_2$ is singular at $r=2$, but $R_0$ is finite there. 
Since $R_0\geq 0$ must hold, 
the allowed range of $b$ is restricted as 
\begin{align}
&b_2\leq b\leq b_1 \ \mathrm{for} \ r> 2,
\\
&b\geq b_2, \ b\leq b_1 \ \mathrm{for} \ r_{\mathrm{H}}<r<2.
\end{align}
We focus on extremum points of $b_{1}$ and $b_2$,
which 
characterize the photon escape conditions. 
The photon orbits staying at the extrema are known 
as the spherical photon orbits~\cite{Teo:2003}. 
Solving equivalent conditions for the extrema, 
$R_0=0$ and $\mathrm{d}R_0/\mathrm{d}r=0$, 
we obtain $b$ and $q$ as functions of 
the spherical photon orbit radius, respectively,
\begin{align}
b_{\mathrm{SPO}}(r)&=-\frac{r^3-3r^2+a^2r+a^2}{a(r-1)},
\\
q_{\mathrm{SPO}}(r)&=-\frac{r^3(r^3-6r^2+9r-4a^2)}{a^2(r-1)^2}.
\end{align}
Eliminating $r$ from these expressions, we find the extremum values 
\begin{align}
b_i^{\mathrm{s}}(q)=b_{\mathrm{SPO}}(r_i(q)) ~~ (i=1,2),
\end{align}
where $r=r_i(q)$ ($r_2 \geq r_1$) is 
the real solutions to the equation $q=q_{\mathrm{SPO}}$. 
If $q=0$, 
the radii $r_i$ reduce to those of the unstable photon circular orbits 
\begin{align}
\label{eq:r1c}
&r_1^{\mathrm{c}}=r_1(0)=2+2\cos \left[\:\!
\frac{2}{3}\arccos(a)-\frac{2\pi}{3}
\:\!\right],
\\
\label{eq:r2c}
&r_2^{\mathrm{c}}=r_2(0)=2+2\cos \left[\:\!
\frac{2}{3}\arccos(a)
\:\!\right],
\end{align}
respectively.
The extremum points appear as a pair of $r_i$ 
only in the range $r_1^{\mathrm{c}}\leq r_1\leq r_2\leq r_2^{\mathrm{c}}$,
and the corresponding range of $q$ is $0\leq q\leq 27$.
The pair coincides with each other at $r_1=r_2=3$, where $q=27$.

\begin{figure}[t]
\centering
 \includegraphics[width=8.0cm,clip]{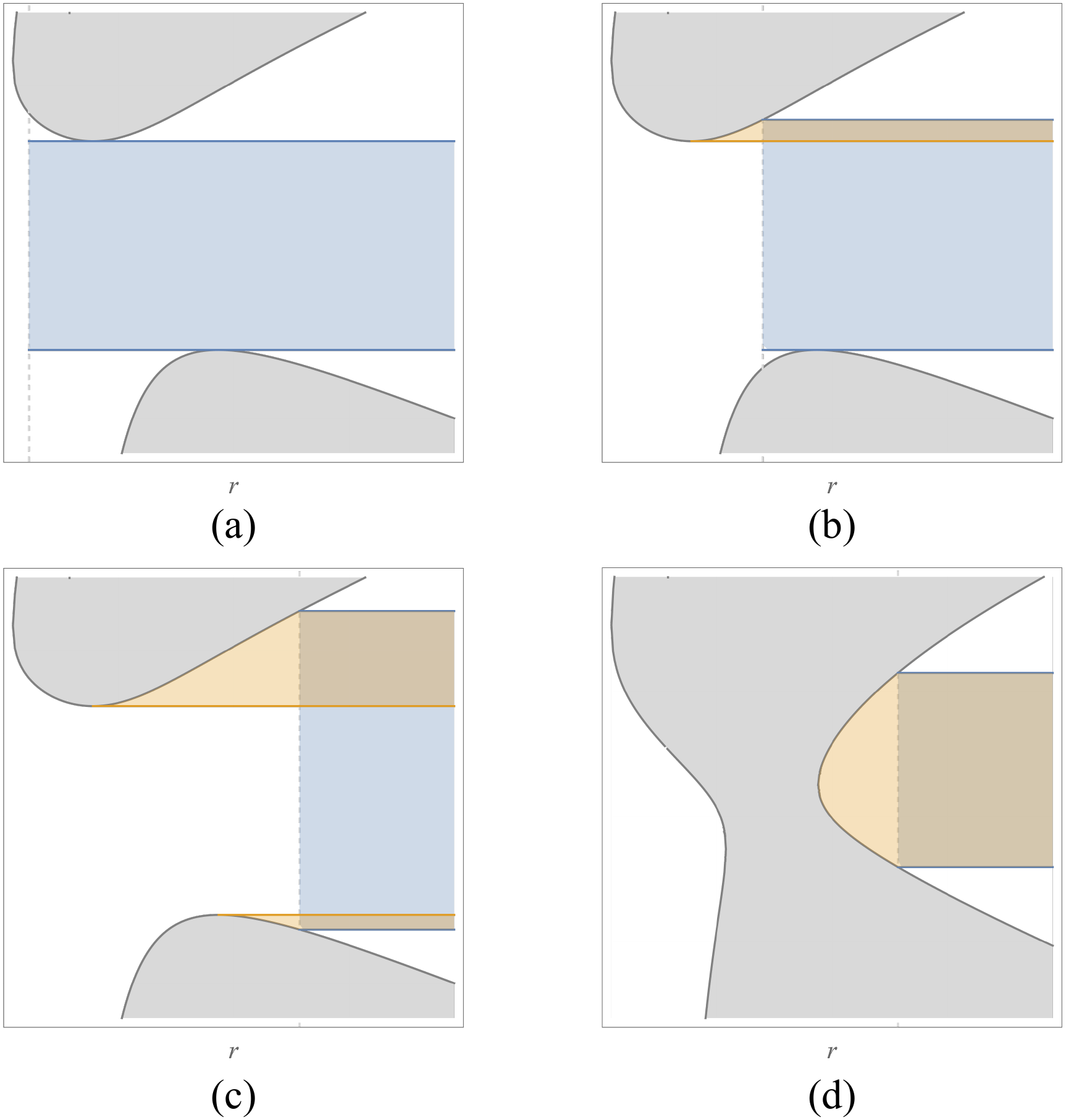}
 \caption{
Typical shape of allowed region for the radial photon motion 
and the parameter range of $b$ for escape photons. 
Gray curves denote $b_i$ ($i=1, 2$) 
as a function of $r$ for fixed $q$,
and gray shaded regions denote forbidden regions, 
while other regions denote allowed regions.
Vertical dashed lines are the radial coordinate value of 
the emission point $r=r_*$. 
The functions $b_i$ in (a)--(c) have extrema $b_i^{\mathrm{s}}$ at $r=r_i$, 
respectively, and are typical shape for $0\leq q<27$, 
where (a) $r_{\mathrm{H}}<r_*<r_1$, (b) $r_1\leq r_* <r_2$, 
and (c) $r_*\geq r_2$. 
The functions $b_i$ in (d) are typical shape for $q>27$, 
in which the allowed region is divided into two disconnected ones, 
and no longer have extrema. 
Blue and orange shaded regions show 
the parameter range of $b$ for an escape photon 
initially with $\sigma_r=1$ and $\sigma_r=-1$, respectively. }
 \label{fig:potentials}
\end{figure}

We can see typical shape of $b_i$ as gray curves 
in Figs.~\ref{fig:potentials}(a)--\ref{fig:potentials}(c) 
for $0\leq q<27$ and in Fig.~\ref{fig:potentials}(d) for $q>27$. 
Gray shaded regions denote forbidden regions of photon motion, 
while other regions denote allowed regions. 
Let us use them to specify whole parameter regions of photon escape 
in a visual manner. 
Let $r_*$ be the radial coordinate value of the emission point, 
which is drawn by dashed vertical lines in the figures. 
In the case $r_{\mathrm{H}}<r_*<r_1$, 
only photons initially emitted outward (i.e., $\sigma_r=1$) 
with $b_2^{\mathrm{s}}<b<b_1^{\mathrm{s}}$ can escape 
[see blue shaded region in Fig.~\ref{fig:potentials}(a)]. 
In the case $r_1\leq r_*<r_2$, 
photons initially emitted outward (i.e., $\sigma_r=1$) 
with $b_2^{\mathrm{s}}<b\leq b_1(r_*; q)$ can escape 
[see blue shaded region in Fig.~\ref{fig:potentials}(b)], and 
photons initially emitted inward (i.e., $\sigma_r=-1$) 
with $b_1^{\mathrm{s}}<b<b_1(r_*; q)$ also can escape 
[see orange shaded region in Fig.~\ref{fig:potentials}(b)]. 
In the case $r_*\geq r_2$, 
photons initially emitted outward (i.e., $\sigma_r=1$) 
with $b_2(r_*; q)\leq b \leq b_1(r_*; q)$ can escape 
[see blue shaded region in Fig.~\ref{fig:potentials}(c)], and 
photons initially emitted inward (i.e., $\sigma_r=-1$) 
with $b_1^{\mathrm{s}}<b<b_1(r_*; q)$ 
or $b_2(r_*; q)<b<b_2^{\mathrm{s}}$ also can escape 
[see orange shaded region in Fig.~\ref{fig:potentials}(c)]. 
For $q\geq 27$, note that the allowed region is disconnected. 
Therefore, if $r_*$ takes a value in the outer allowed region, then 
photons must have $b_2(r_*; q)\leq b\leq b_1(r_*; q)$ 
and always can escape 
[see blue and orange shaded region in Fig.~\ref{fig:potentials}(d)].

Let us summarize the photon escape conditions 
as a region in the $(b, q)$ parameter space for a fixed $r_*$.
To complete it, we introduce the following four ranges of $r_*$:
\begin{align}
\mathrm{(i)}~&r_{\mathrm{H}}<r_*<r_1^{\mathrm{c}},
\\
\mathrm{(\ii)}~&r_1^{\mathrm{c}} \leq r_* <3,
\\
\mathrm{(\iii)}~&3\leq r_*<r_2^{\mathrm{c}},
\\
\mathrm{(\iv)} ~&r_2^{\mathrm{c}}\leq r_*.
\end{align}
To identify the photon escape parameter regions completely, 
we introduce two specific values of $q$, 
\begin{align}
&q_*=q_{\mathrm{SPO}}(r_*),
\\
&q_{\max}= \frac{r_*^3}{r_*-2}.
\end{align}
For $q=r^3/(r-2)$, the functions $b_{1,2}$ are degenerate, 
and for $q>r^3/(r-2)$, the position $r$ enters the forbidden region. 
Therefore, the maximum value of $q$ is limited 
by $q_{\max}$ for the photon emission.
Finally, we can summarize the photon escape parameter region 
in the $(b, q)$ plane as Table~\ref{table:escapecond}.
Though the photon escape regions shown in Refs.~\cite{Ogasawara:2019mir, Ogasawara:2020frt} 
were only Cases~(i) and (\ii), 
here we have identified them for all the ranges of $r_*$. 
\begin{table}
\caption{Photon escape parameter region in the Kerr spacetime.}
\begin{tabular}{llll}
\hline\hline
Case&$q$&$b ~(\sigma_r=+)$&$b~ (\sigma_r=-)$
\\
\hline
(i) $r_{\mathrm{H}}<r_*<r_1^{\mathrm{c}}$~~
&$0\leq q\leq 27$
&$b_2^{\mathrm{s}}<b<b_1^{\mathrm{s}}$
&n/a
\\
\hline
(\ii) $r_1^{\mathrm{c}}\leq r_*<3$
&$0\leq q<q_*$
&$b_2^{\mathrm{s}}<b\leq b_1(r_*; q)$
&$b_1^{\mathrm{s}}<b<b_1(r_*; q)$
\\
&$q_*\leq q\leq 27$
&$b_2^{\mathrm{s}}<b<b_1^{\mathrm{s}}$
&n/a
\\
\hline
(\iii) $3\leq r_*<r_2^{\mathrm{c}}$
&$0\leq q <q_*$
&$b_2^{\mathrm{s}}<b\leq b_1(r_*; q)$
&$b_1^{\mathrm{s}}<b<b_1(r_*; q)$
\\
&$q_*\leq q<27$
&$b_2(r_*;q)\leq b\leq b_1(r_*;q)$~~
&$b_2(r_*;q)<b<b_2^{\mathrm{s}}$, $b_1^{\mathrm{s}}<b<b_1(r_*;q)$ 
\\
&$27\leq q\leq q_{\max}$~~
&$b_2(r_*;q)\leq b\leq b_1(r_*;q)$
&$b_2(r_*;q)<b<b_1(r_*;q)$
\\
\hline
(\iv) $r_2^{\mathrm{c}} \leq r_*$
&$0\leq q <27$
&$b_2(r_*;q)\leq b\leq b_1(r_*;q)$
&$b_2(r_*;q)<b<b_2^{\mathrm{s}}$, $b_1^{\mathrm{s}}<b<b_1(r_*;q)$
\\
&$27\leq q\leq q_{\max}$
&$b_2(r_*;q)\leq b\leq b_1(r_*;q)$
&$b_2(r_*;q)< b< b_1(r_*;q)$
\\
\hline\hline
\end{tabular}
\label{table:escapecond}
\end{table}

\section{Dynamics of an emitter}
\label{sec:4}
\begin{figure}[t]
\centering
 \includegraphics[width=10cm,clip]{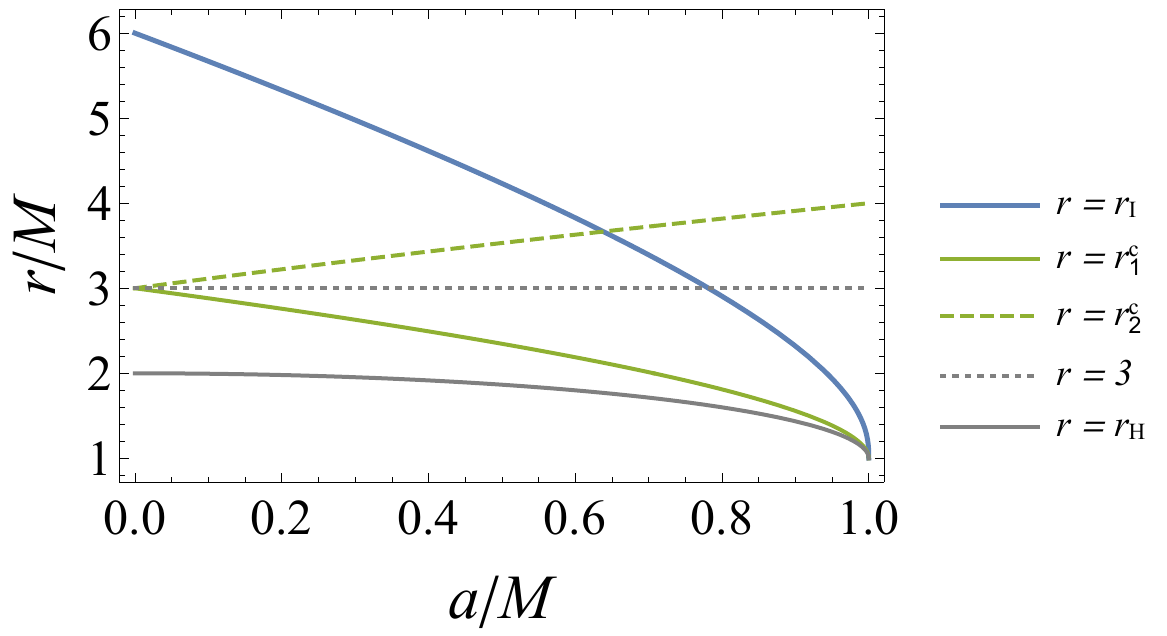}
 \caption{Characteristic radii of particle circular orbits as functions of $a$. 
Blue solid curve: the ISCO radius given by Eq.~\eqref{eq:rI}. 
Green solid curve: the radius of photon circular prograde orbit given by Eq.~\eqref{eq:r1c}.
Green dashed curve: the radius of photon circular retrograde orbit given by Eq.~\eqref{eq:r2c}. 
Gray dotted line is $r=3$, and 
gray solid curve denotes the horizon radius. 
The equality $r_{\mathrm{I}}=r_2^{\mathrm{c}}$ holds at $a=0.6382\ldots$, and 
$r_{\mathrm{I}}=3$ holds at $a=0.7818\ldots$.}
 \label{fig:radii}
\end{figure}

We consider an emitter ($\kappa=1$) 
freely falling from the ISCO into the horizon. 
Assume that the emitter moves on $\theta=\pi/2$ and 
has the same energy and angular momentum 
as those of a particle orbiting the ISCO. 
Then, the constants $Q$, $E$, and $L$ of the emitter are 
\begin{align}
\label{eq:EI}
Q&=0,
\\
E&=\frac{r_{\mathrm{I}}^{3/2}-2r_{\mathrm{I}}^{1/2}+a}{r_{\mathrm{I}}^{3/4}(r_{\mathrm{I}}^{3/2}-3r_{\mathrm{I}}^{1/2}+2a)^{1/2}},
\\
\label{eq:LI}
L&=\frac{r_{\mathrm{I}}^2-2ar_{\mathrm{I}}^{1/2}+a^2}{r_{\mathrm{I}}^{3/4}(r_{\mathrm{I}}^{3/2}-3r_{\mathrm{I}}^{1/2}+2a)^{1/2}},
\end{align}
where $r_{\mathrm{I}}$ is the radius of the ISCO~\cite{Bardeen:1972fi}, 
\begin{align}
\label{eq:rI}
r_{\mathrm{I}}&=3+Z_2-\left[\:\!
(3-Z_1)(3+Z_1+2Z_2)
\:\!\right]^{1/2},
\\
Z_1&=1+(1-a^2)^{1/3}\left[\:\!
(1+a)^{1/3}+(1-a)^{1/3}
\:\!\right],
\\
Z_2&=(3a^2+Z_1^2)^{1/2}.
\end{align}
Figure~\ref{fig:radii} shows the relation between $r_{\mathrm{I}}$ 
and the characteristic radii appearing in Table~\ref{table:escapecond}. 
A blue solid curve shows $r=r_{\mathrm{I}}$. 
Green solid and dashed curves
are $r=r_1^{\mathrm{c}}$ and $r=r_2^{\mathrm{c}}$, respectively. 
A gray dotted line denotes $r=3$. 
Note that $r_{\mathrm{I}} > r_1^{\mathrm{c}}$ holds for all the range of $a$, 
and $r_{\mathrm{I}}>r_2^{\mathrm{c}}$ holds only for $a<0.6382\ldots$, and 
$r_{\mathrm{I}}>3$ holds only for $a<0.7818\ldots$.
Therefore, when we consider a plunge orbit from the ISCO, 
all the cases in Table~\ref{table:escapecond} appear only for $a< 0.6382\ldots$; 
on the other hand, only Cases~(i)--(\iii) appear for $0.6382\ldots\leq a< 0.7818\ldots$, 
and only Cases~(i)--(\ii) appear for $0.7818\ldots\leq a < 1$. 
The Schwarzschild case $a=0$ is special because $r_i^{\mathrm{c}}=3$ 
and will be discussed in Appendix~\ref{sec:A}.

\begin{figure}[t]
\centering
 \includegraphics[width=7.9cm,clip]{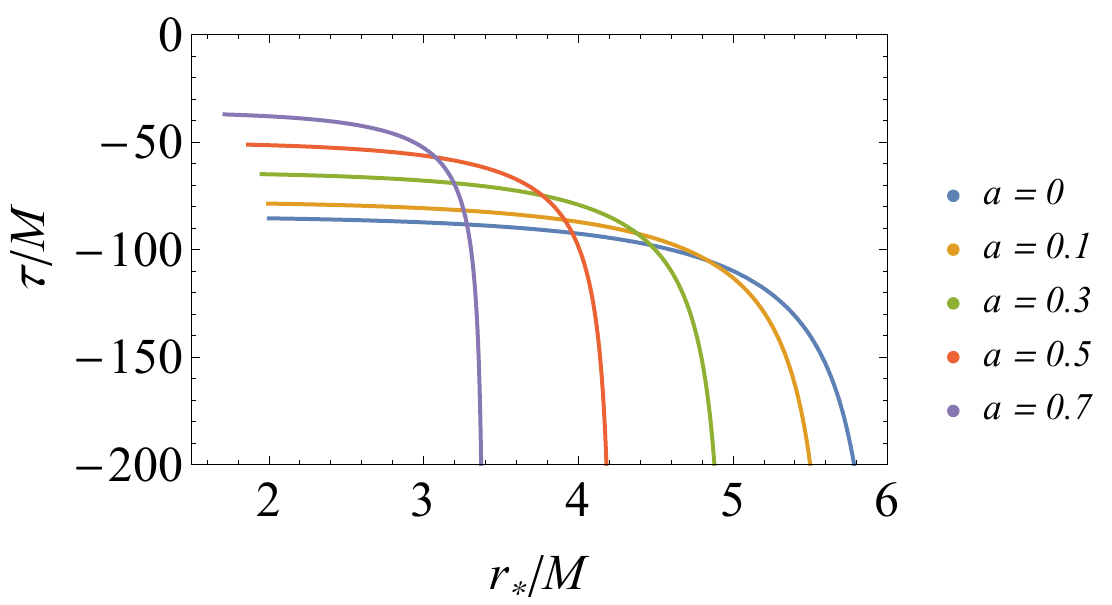}
 \includegraphics[width=8.4cm,clip]{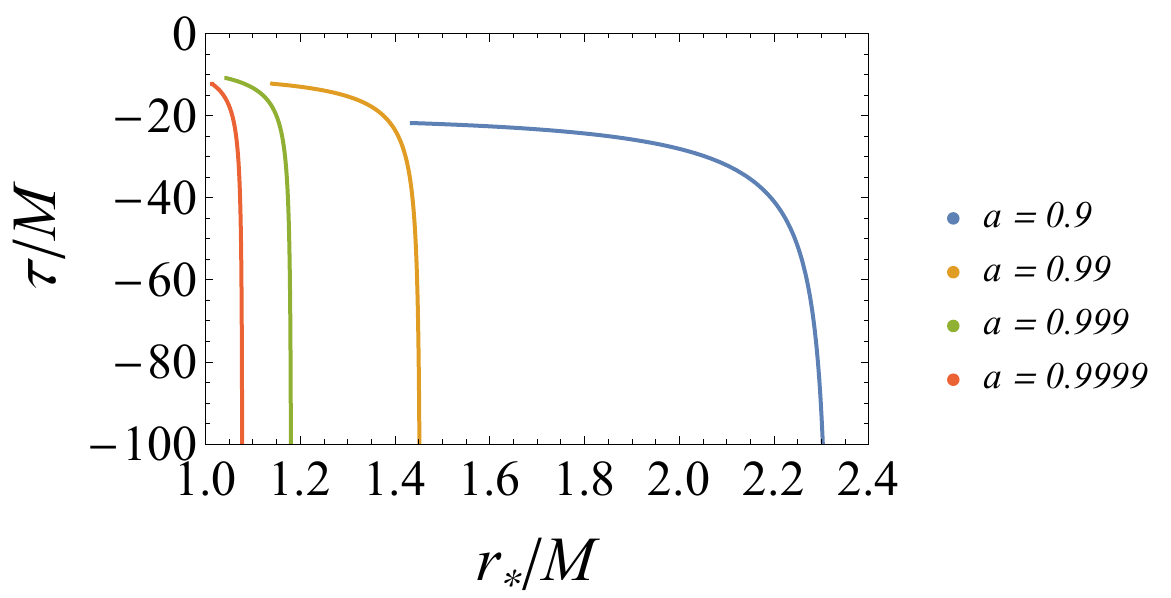}
 \caption{Radial motion of the emitter for each value of $a$. 
 It falls from the vicinity of the ISCO into the horizon. 
 Each curve terminates at the horizon radius.}
 \label{fig:taur0109}
\end{figure}

\begin{figure}[t]
\centering
 \includegraphics[width=5.9cm,clip]{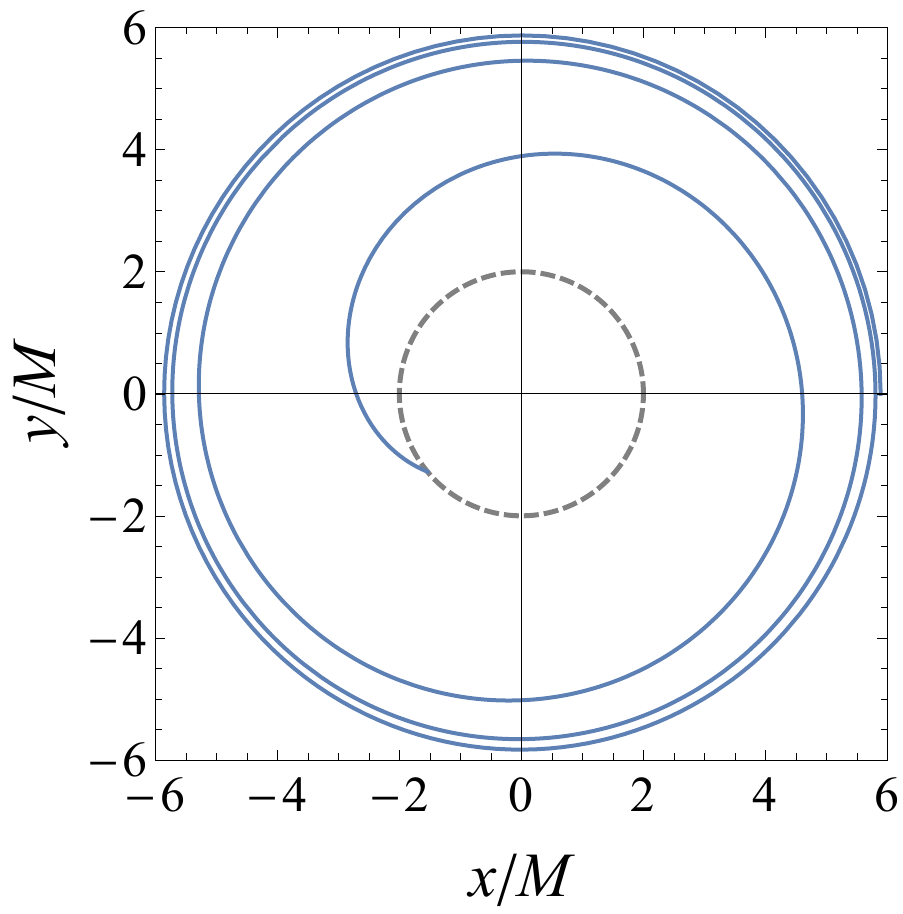}
 ~~~~~~
 \includegraphics[width=5.9cm,clip]{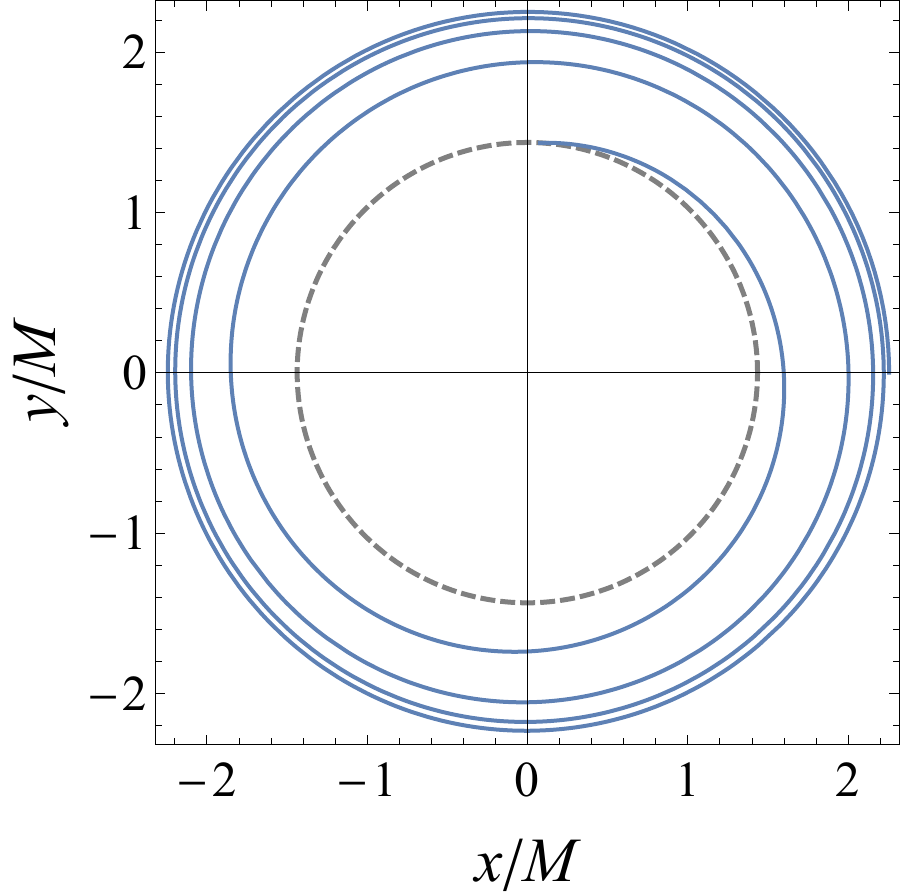}
 \caption{
Top view of the plunge trajectories of an emitter 
from the vicinity of the ISCO into the horizon. 
The blue solid curves show the plunge orbits, 
and the gray dashed circles denote the horizon radii. 
The left figure shows the case $a=0$, for which $r_{\mathrm{H}}= 2$ 
and $r_{\mathrm{I}}=6$, 
and the initial condition is chosen as $\epsilon=0.1118$.
The right figure shows the case $a=0.9$, 
for which $r_{\mathrm{H}}\simeq 1.4358$ and $r_{\mathrm{I}}\simeq 2.3208$, and 
the initial condition is chosen as $\epsilon=0.06379$.
 The size of each frame coincides with each ISCO radius. 
The coordinates are defined by $(x, y)=(r \cos \psi, r\sin \psi)$, 
where $\psi$ is the angle in the ingoing Kerr-Schild coordinates (see Appendix~\ref{sec:B}).}
 \label{fig:orbit09}
\end{figure}

The equations of motion~\eqref{eq:tdot}--\eqref{eq:rdot} for the emitter 
are reduced to 
\begin{align}
\label{eq:tdotpl}
\dot{t}&=\frac{E\:\!(1+a^2 u^2)-2a (L-a E)u^3 
}{a^2 (u-u_+)(u-u_-)},
\\
\label{eq:phidotpl}
\dot{\varphi}&=\frac{
L \:\!u^2-2(L-a E) u^3
}{a^2 (u-u_+)(u-u_-)},
\\
\label{eq:udot}
\dot{u}&=\sqrt{2} \:\!(L-a E) \:\!u^2 (u-u_{\mathrm{I}})^{3/2},
\end{align}
where $\sigma_r=-1$, and we have introduced a new variable $u=r^{-1}$ 
and constants
$u_{+}=r_{\mathrm{H}}^{-1}$, 
$u_{-}=r_{\mathrm{-}}^{-1}=(1-\sqrt{1-a^2})^{-1}$,
and $u_{\mathrm{I}}=r_{\mathrm{I}}^{-1}$. 
Solving these equations, we obtain 
\begin{align}
\label{eq:taur}
\tau-\tau_0&=-\frac{\sqrt{2}}{(L-a E) \:\!u_{\mathrm{I}}^2\sqrt{u-u_{\mathrm{I}}}} {}_2 F_1 \left(-\frac{1}{2}, 2; \frac{1}{2}; 1-\frac{u}{u_{\mathrm{I}}}\right),
\\
\label{eq:tr}
t-t_0&=\frac{1}{\sqrt{2} \:\!a^2(L-a E)}\bigg[\!
-\frac{4\left[\:\!
E-a \:\!(L-a E)\:\! u_{\mathrm{I}}^2
\:\!\right]}{u_{\mathrm{I}}(u_+-u_{\mathrm{I}})(u_{-}-u_{\mathrm{I}})\sqrt{u-u_{\mathrm{I}}}}
+\frac{a^2 E (u_{\mathrm{I}}-3u)}{u_{\mathrm{I}}^2u \sqrt{u-u_{\mathrm{I}}}}
\cr
&~~~~~~~~~~~~~~~~~~~~~~~~~
-\frac{a^2 E (3+4 u_{\mathrm{I}})}{u_{\mathrm{I}}^{5/2}}\tan^{-1}\Big(
\sqrt{\frac{u-u_{\mathrm{I}}}{u_{\mathrm{I}}}}\Big)
\cr
&~~~~~~~~~~~~~~~~~~~~~~~~~
+\frac{4 (a L u_{-}-2E)}{(u_{-}-u_{+})(u_{-}-u_{\mathrm{I}})^{3/2}} \tanh^{-1}\Big(
\sqrt{\frac{u-u_{\mathrm{I}}}{u_{-}-u_{\mathrm{I}}}}
\Big)
\cr
&~~~~~~~~~~~~~~~~~~~~~~~~~
-\frac{4 (a L u_{\mathrm{+}}-2 E)}{(u_{-}-u_{+})(u_{+}-u_{\mathrm{I}})^{3/2}}
\tanh^{-1} \Big(\sqrt{\frac{u-u_{\mathrm{I}}}{u_{+}-u_{\mathrm{I}}}}\Big)
\:\!\bigg],
\\
\label{eq:phir}
\varphi-\varphi_0&=\frac{\sqrt{2}}{a^2 (L-a E)}\bigg[\!
-\frac{L-2\:\! (L-a E)\:\! u_{\mathrm{I}}}{(u_+-u_{\mathrm{I}})(u_{-}-u_{\mathrm{I}})\sqrt{u-u_{\mathrm{I}}}}
\cr
&~~~~~~~~~~~~~~~~~~~~~
+
\frac{L-2\:\!(L-a E)\:\!u_{+}}{(u_{-}-u_{+})(u_{+}-u_{\mathrm{I}})^{3/2}}\tanh^{-1}\Big(
\sqrt{\frac{u-u_{\mathrm{I}}}{u_{+}-u_{\mathrm{I}}}}
\Big)
\cr
&~~~~~~~~~~~~~~~~~~~~~
-\frac{L-2\:\!(L-a E)\:\!u_-}{(u_{-}-u_{+})(u_{-}-u_{\mathrm{I}})^{3/2}}\tanh^{-1}\Big(
\sqrt{\frac{u-u_{\mathrm{I}}}{u_{-}-u_{\mathrm{I}}}}
\Big)
\bigg],
\end{align}
where $\tau$ is proper time, and ${}_2 F_1$ is the Gaussian hypergeometric function, 
and $\tau_0$, $t_0$, and $\varphi_0$ are arbitrary constants and are set to zero in the following. 
Further assume that the radial position of the emitter, 
$r_*(\tau)$, at initial time $\tau=\tau_{\mathrm{i}}$ is given by
\begin{align}
r_*(\tau_{\mathrm{i}})=r_{\mathrm{I}}-\epsilon,
\end{align}
where $0<\epsilon\ll 1$.%
\footnote{In the limit $\epsilon \to 0$, 
then $\tau_{\mathrm{i}}\to -\infty$.} 
It is worth noting that 
the plunge orbit does not depend on the choice of $\epsilon$; in other words, 
it is unique up to the gauge freedom of $\tau$, $t$, and $\varphi$.
The initial radial velocity $\dot{r}_*(\tau_{\mathrm{i}})$ is determined through Eq.~\eqref{eq:udot}.
The change of $r_*(\tau)$ in Eq.~\eqref{eq:taur} is shown in 
Fig.~\ref{fig:taur0109}. 
Each curve gradually decreases as $\tau$ increases, 
and it eventually terminates at the horizon, 
$r_*(\tau_{\mathrm{f}})=r_{\mathrm{H}}$, which means that 
the time duration $\tau_{\mathrm{f}}-\tau_{\mathrm{i}}$ is always finite.
As seen in Eq.~\eqref{eq:tr}, however, the corresponding coordinate time duration 
is infinite because of the infinite gravitational redshift at the horizon. 
Typically, the emitter leaving from the vicinity of the ISCO will orbit 
with little change in its orbital radius for a while, i.e., 
the radial velocity component will be relatively small in an early phase. 
After that, the radial velocity will gradually increase 
and eventually shift into a plunging orbit toward the horizon.
Figure~\ref{fig:orbit09} shows examples of the plunge orbits for $a=0$ and $a=0.9$. 

\section{Photon escape probability}
\label{sec:5}
To evaluate the escape probability of a photon 
isotropically emitted from the emitter, 
we introduce 
a tetrad frame 
at rest with respect to the emitter, 
$\zeta^{(\mu)}_*=\zeta^{(\mu)}|_{r=r_*,\theta=\theta_*}$, 
where
\begin{align}
\zeta^{(0)}&=-E\:\!\mathrm{d}t-\frac{\sqrt{R_1}}{\Delta}\:\!\mathrm{d}r+L\mathrm{d}\varphi,
\\
\zeta^{(1)}&=\frac{1}{\sqrt{\rho^2-r^2 \Delta A}} \left[\:\!
-2a \sqrt{R_1} \:\!\mathrm{d}t+L r^3\:\!\mathrm{d}r+\frac{A}{r}\sqrt{R_1}\:\!\mathrm{d}\varphi
\:\!\right],
\\
\zeta^{(2)}&= r\:\!\mathrm{d}\theta,
\\
\zeta^{(3)}&=\frac{1}{\sqrt{\rho^2-r^2 \Delta A}} \left[\:\!
(r^2\Delta- E \rho)\:\!\mathrm{d}t-\frac{\rho\sqrt{R_1}}{\Delta}\:\!\mathrm{d}r
+L \rho\:\!\mathrm{d}\varphi
\:\!\right],
\end{align}
where $\zeta^{(0)}_*$ coincides with the momentum of the emitter, 
and 
\begin{align}
\rho&=E A -2aL r.
\end{align}
The tetrad components of the photon momentum in this frame are 
$k^{(\mu)}_*=k^{(\mu)}|_{r=r_*, \theta=\theta_*}$, where 
$k^{(\mu)}=k^{a}\zeta^{(\mu)}_a$ are given by
\begin{align}
k^{(0)} &=\frac{
b \:\!r[\:\!
2a E+L (r-2)
\:\!]-\rho-\sigma_r \sqrt{R_0 R_1}
}{r^2\Delta},
\\
k^{(1)}&=\frac{r(\sigma_r L \sqrt{R_0}+b \sqrt{R_1})}{\sqrt{\rho^2-r^2 \Delta A}},
\\
k^{(2)}&=\sigma_\theta \frac{\sqrt{q}}{r},
\\
k^{(3)}
&=\frac{L\:\! r \rho \left[\:\!
2a+b(r-2)
\:\!\right]-(A-2a\:\!b\:\!r)(E \rho-r^2\Delta )-\sigma_r \rho \sqrt{R_0R_1}}{r^2\Delta \sqrt{\rho^2-r^2\Delta A}}.
\end{align}
Note that $\sigma_r$ and $\sigma_\theta$ are for an emitted photon. 
\begin{figure}[t]
\centering
 \includegraphics[width=5cm,clip]{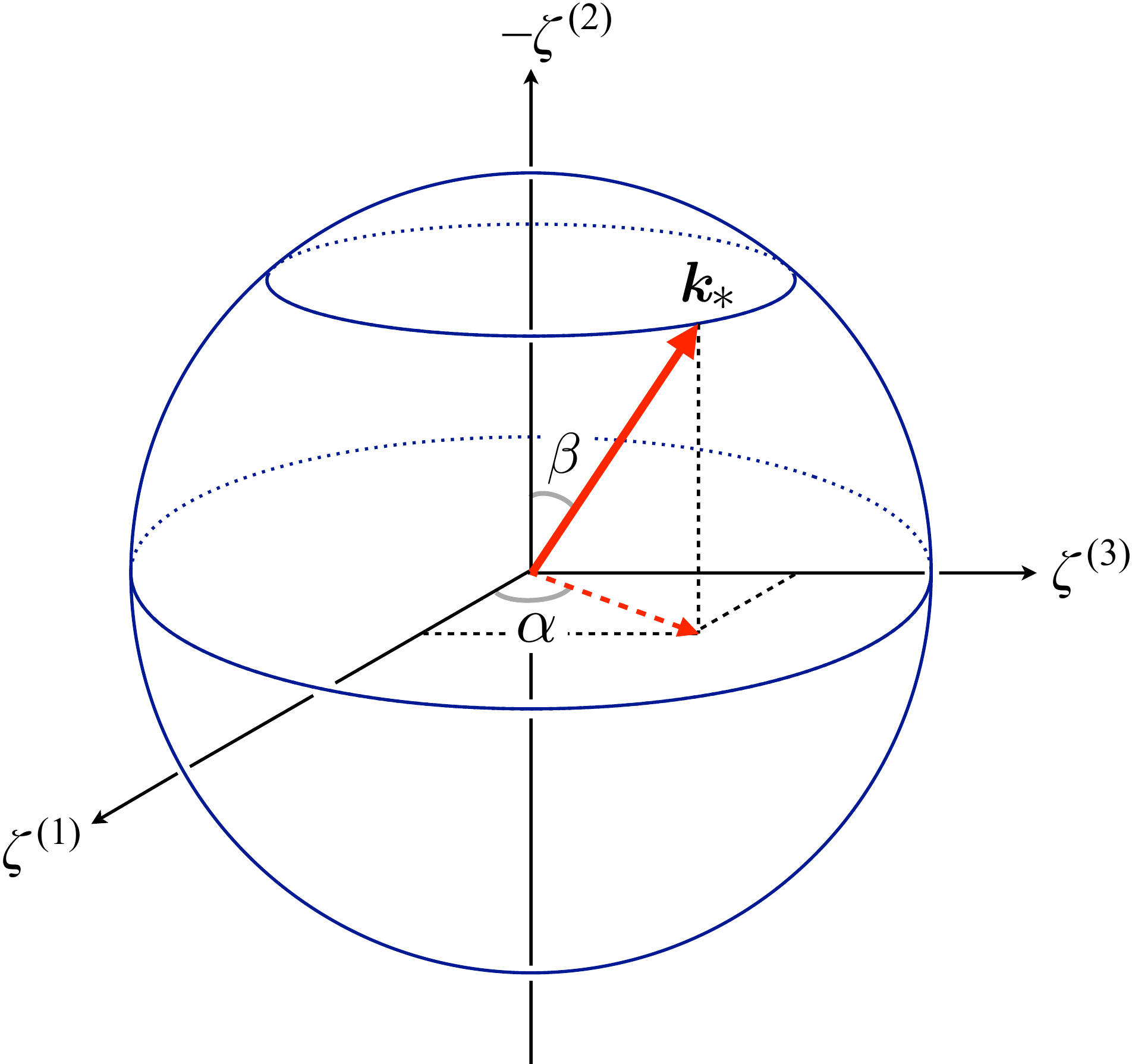}
 \caption{Emission angles~$(\alpha, \beta)$ defined at the rest frame of the emitter.}
 \label{fig:angles}
\end{figure}
We can relate the impact parameters $(b,q)$ to the emission angle $(\alpha, \beta)$ 
measured in the rest frame by
\begin{align}
\cos\alpha &=\frac{k_*^{(1)}}{\sqrt{\big(k_*^{(1)}\big)^2+\big(k_*^{(3)}\big)^2}},
\quad
\sin \alpha =\frac{k_*^{(3)}}{\sqrt{\big(k_*^{(1)}\big)^2+\big(k_*^{(3)}\big)^2}},
\\
\cos \beta&=\frac{k_*^{(2)}}{k_*^{(0)}},
\quad
\sin\beta =-\frac{\sqrt{\big(k_*^{(1)}\big)^2+\big(k_*^{(3)}\big)^2}}{k_*^{(0)}},
\end{align}
where the sign of $\cos \beta$ flips only with the sign flip of $\sigma_\theta$. 
Figure~\ref{fig:angles} illustrates the relation between $k_*^{(\mu)}$ and $(\alpha, \beta)$.
Then, we define the photon escape probability as
\begin{align}
\label{eq:P}
P(r_*; a)=\frac{1}{4\pi} \int\!\!\!\!\!\int_{\mathcal{E}} \mathrm{d}\alpha \wedge \mathrm{d}\beta,
\end{align}
where $\mathcal{E}$ denotes the whole parameter region of escape photons 
in the $(\alpha, \beta)$ plane (i.e., the photon escape 
cone~\cite{Synge:1966okc,Semerak:1996,Ogasawara:2019mir}).
The region $\mathcal{E}$ are translated from the photon escape parameter region 
in Table~\ref{table:escapecond}.
One of the practical ways to evaluate $P$ is shown in Appendix~\ref{sec:C}.

\begin{figure}[t]
\centering
 \includegraphics[width=7.8cm,clip]{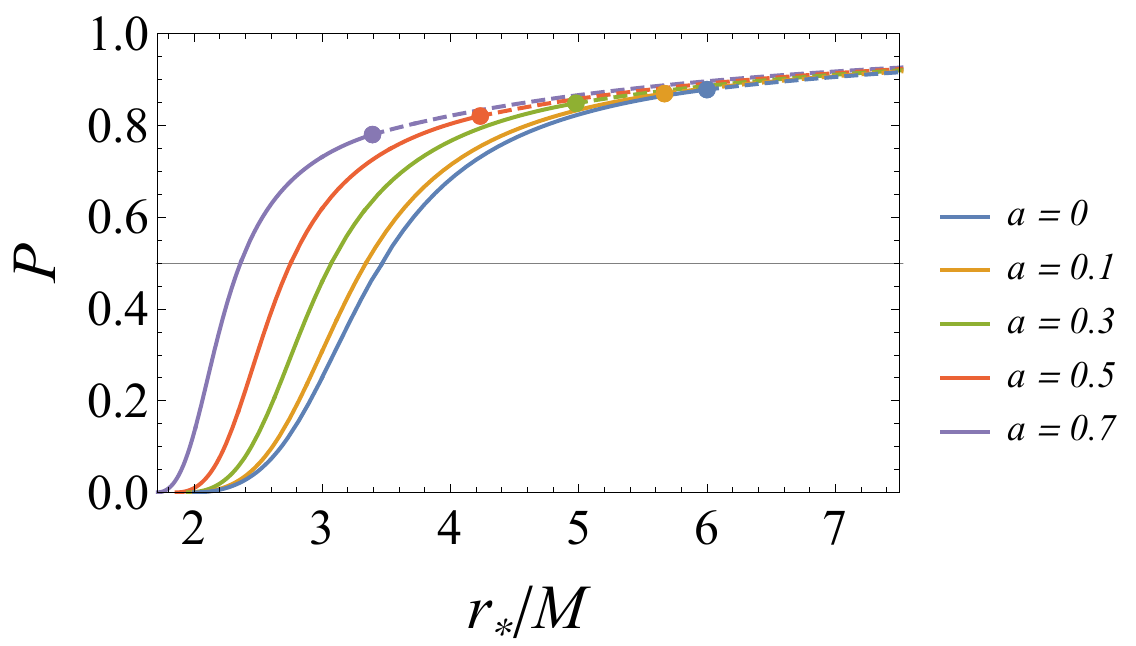}
 \includegraphics[width=8.4cm,clip]{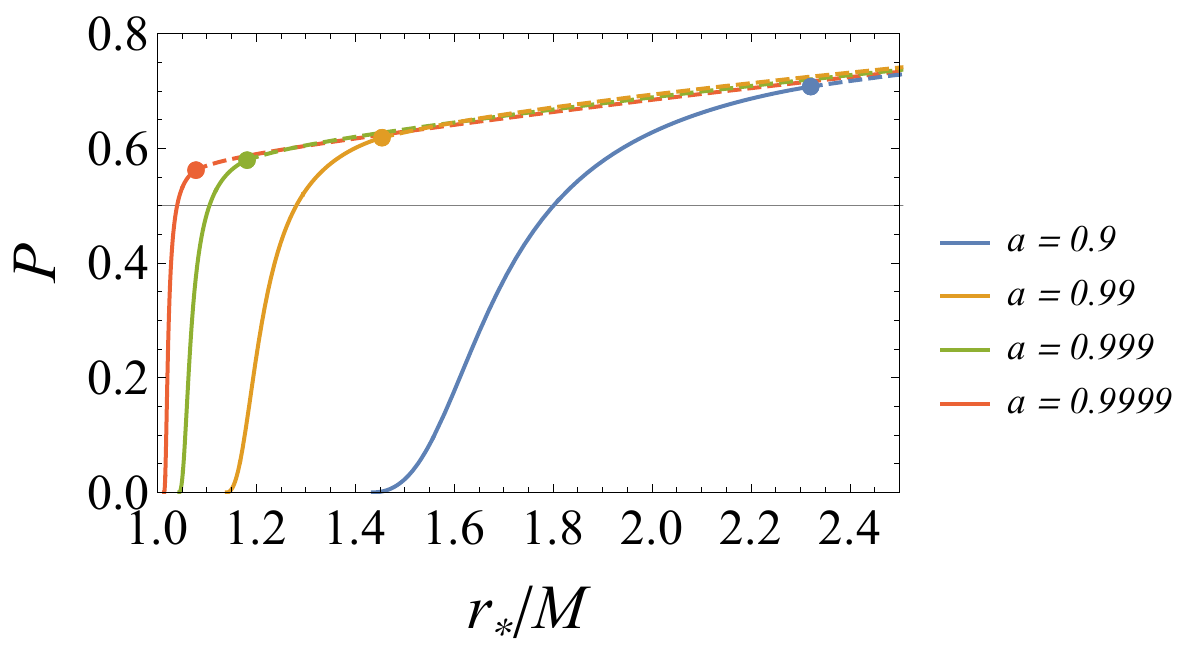}
 \caption{
 Photon escape probability $P(r_*; a)$ for each value of $a$. 
 Solid curves: the escape probabilities of a photon emitted from 
 the emitter plunging from the ISCO into the horizon. 
 Dashed curves and dots are the escape probabilities of a photon emitted 
 from an emitter on circular geodesics and the ISCO, respectively.}
 \label{fig:a0109}
\end{figure}

Figure~\ref{fig:a0109} shows $P$ of $r_*$ for a fixed value of $a$, 
which are shown by solid curves. 
Each dashed curve shows the escape probability of a photon emitted 
from an emitter orbiting circular geodesics~\cite{Igata:2019hkz}. 
Each dot at the connection between solid and dashed curves shows 
the limit value $P(r_{\mathrm{I}};a)$ 
in the limit $r_*\to r_{\mathrm{I}}$ (i.e., $\epsilon\to 0$).%
\footnote{The value $P(r_{\mathrm{I}};a)$ decreases as $a$ increases 
but approaches to a finite value $P= 0.5464\ldots$
even in the extremal limit $a\to 1$~\cite{Igata:2019hkz,Gates:2020els}.} 
We find that once the emitter gently starts to fall from the vicinity of the ISCO, 
the probability $P$ decreases monotonically with $r_*$. 
Even when the emitter crosses the unstable photon circular orbit radii, 
$P$ shows no characteristic behavior.
As the left endpoint of each curve shows, 
$P$ vanishes when the emitter reaches the horizon.

\begin{figure}[t]
\centering
 \includegraphics[width=8cm,clip]{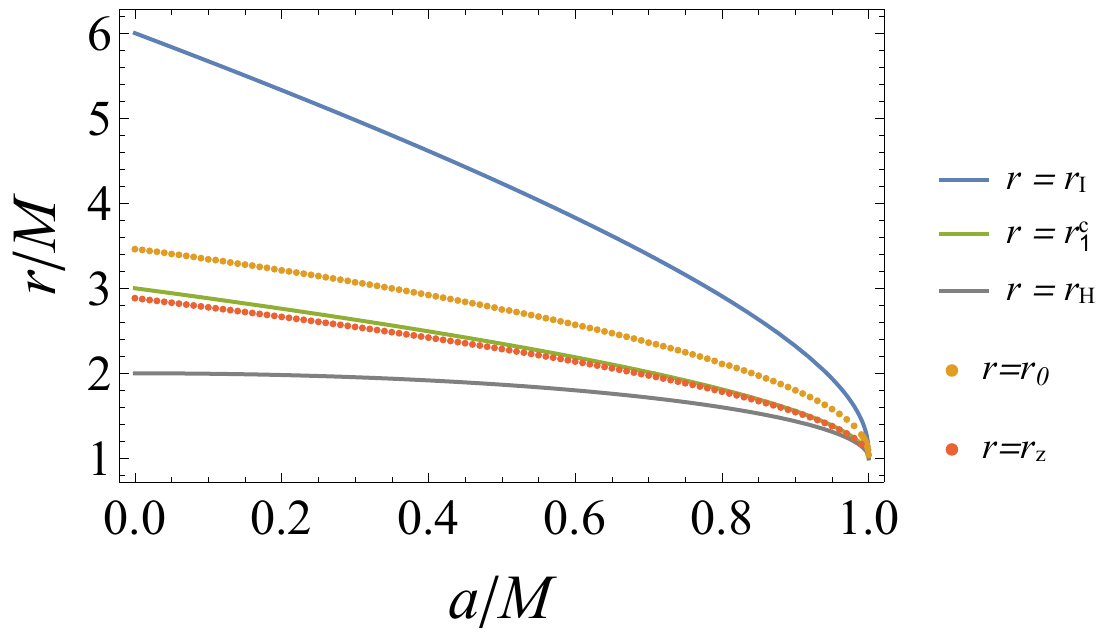}
 \caption{Plots of $r_0$ and $r_{\mathrm{z}}$. Orange dots: $r=r_0$. 
Red dots: $r=r_{\mathrm{z}}$. 
Blue solid curve: the ISCO radius given by Eq.~\eqref{eq:rI}. 
Green solid curve: the radius of photon circular prograde orbit given by Eq.~\eqref{eq:r1c}.
Gray solid curve: the horizon radius. 
}
 \label{fig:half}
\end{figure}

\begin{figure}[t]
\centering
 \includegraphics[width=16.0cm,clip]{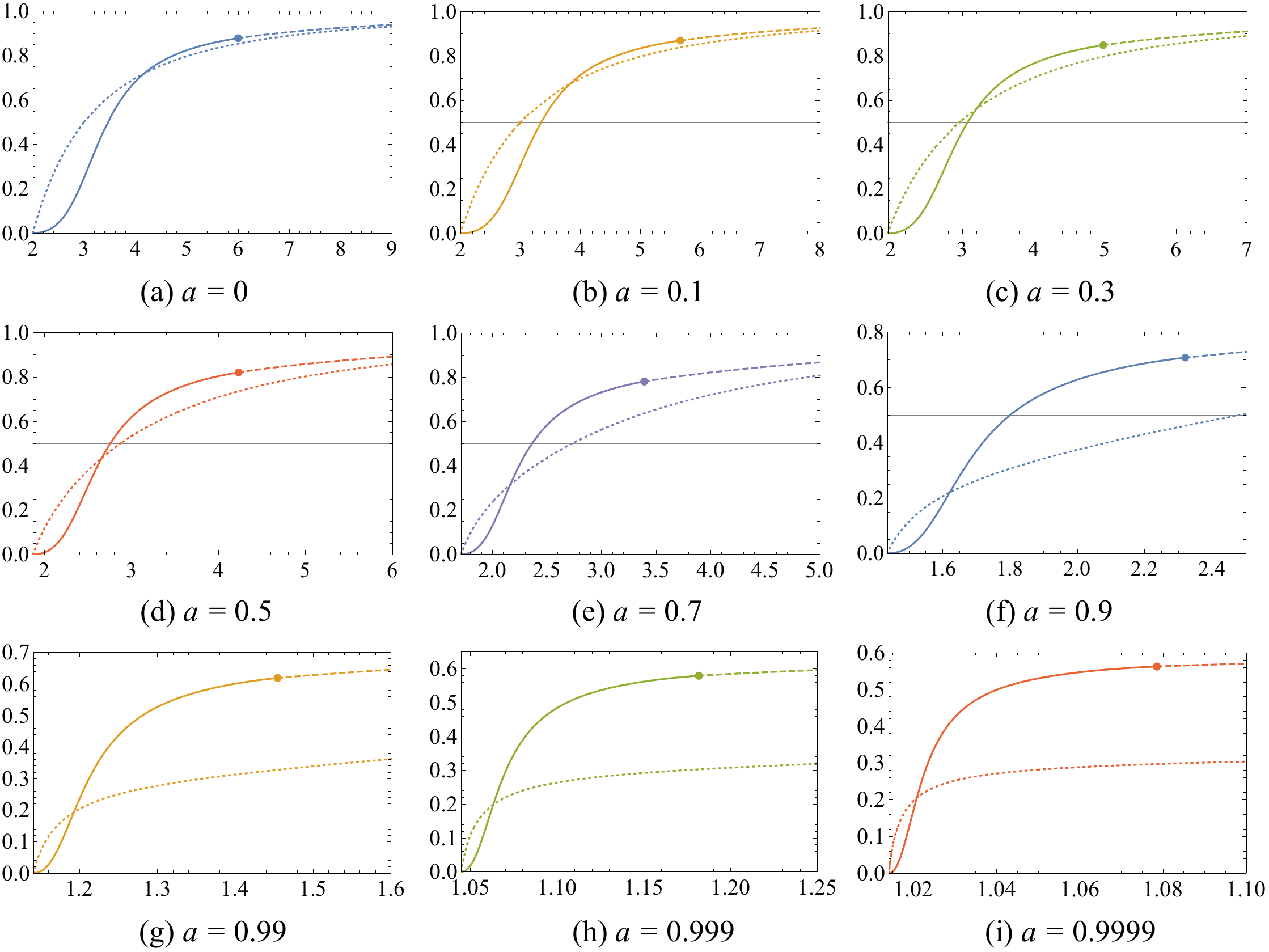}
 \caption{
Escape probabilities of a photon emitted from an emitter in various states of motion. 
Solid curves denote $P$, and dotted curves denote $P_{\mathrm{L}}$, 
as functions of $r_*$.
Dashed curves: the escape probabilities of a photon emitted from an emitter on circular geodesics.}
 \label{fig:LNRF}
\end{figure}

As the emitter moves along the plunge orbit, 
$P$ will certainly be less than 
a half somewhere along the way (see gray lines in Fig.~\ref{fig:a0109}).
Let $r_0$ be the radius at which $P=1/2$. 
For $a=0$, we find that $r_0 \simeq 3.46$, 
which is outside the photon sphere at $r=3$. 
Since the photon escape probability 
for a static light source is just a half at $r=3$~\cite{Synge:1966okc}, 
our result shows that $P$ tends to be suppressed 
by the effect of the proper motion [see also Fig.~\ref{fig:LNRF}(a)]. 
Several values of $r_0$ are summarized in Table~\ref{table:rs}, 
and the plots of $r_0$ are shown in Fig.~\ref{fig:half} (see orange dots). 
The radius $r_0$ decreases monotonically as $a$ increases.
We can roughly evaluate $r_0$ as being 
approximately intermediate value between $r_{\mathrm{I}}$ and $r_{\mathrm{H}}$,
\begin{align}
\frac{2\:\! r_{0}}{r_{\mathrm{I}}+r_{\mathrm{H}}}\simeq 1.
\end{align}
Therefore, 
if we adopt $r_0$ as one of the indicators 
that divides the region where plunge orbits are observable or not, 
we can conclude that our plunge orbits 
are observable up to the middle of $r_{\mathrm{I}}$ and $r_{\mathrm{H}}$.
\begin{table}[t]
\caption{Numerical values of characteristic radii for several values of $a$. }
\centering
\begin{tabular}{cccccccccccccc}
\hline\hline
$a$&$0$&$0.1$&$0.2$&$0.3$&$0.4$&$0.5$&$0.6$&$0.7$&$0.8$&$0.9$
&$0.99$&$0.999$&$0.9999$
\\
\hline
$r_{\mathrm{I}}$&$6$&$5.6693$&$5.3294$&$4.9786$&$4.6143$&$4.2330$&$3.8290$&$3.3931$&$2.9066$&$2.3208$
&$1.4544$&$1.1817$&$1.0785$
\\
$r_{0}$&$3.46$&$3.34$&$3.21$&$3.07$&$2.92$&$2.75$&$2.57$&$2.36$&$2.11$&$1.80$&$1.28$&$1.10$&$1.04$
\\
$r_1^{\mathrm{c}}$&$3$&$2.8821$&$2.7591$&$2.6300$&$2.4933$&$2.3472$&$2.1889$&$2.0133$&$1.8110$&$1.5578$
&$1.1676$&$1.0520$&$1.0163$
\\
$r_{\mathrm{z}}$&$2.883$&$2.775$&$2.663$&$2.545$&$2.419$&$2.285$&$2.138$&$1.974$&$1.784$&$1.543$&$1.167$&$1.051$&$1.017$
\\
$r_{\mathrm{H}}$&$2$&$1.9949$&$1.9797$&$1.9539$&$1.9165$&$1.8660$&$1.8$&$1.7141$&$1.6$&$1.4358$
&$1.1410$&$1.0447$&$1.0141$
\\
\hline\hline
\end{tabular}
\label{table:rs}
\end{table}

Similarly for the case $a\neq 0$, it is useful to compare $r_0$ 
with the radius at which the escape probability of a photon emitted 
from an emitter stationary relative to the black hole
[i.e., an emitter at rest in the locally nonrotating frame~(LNRF)] is a half, 
where the LNRF is given by
\begin{align}
e^{(0)}=\sqrt{\frac{\Sigma\Delta}{A}}\:\!\mathrm{d} t,
\quad
e^{(1)}=\sqrt{\frac{\Sigma}{\Delta}}\:\!\mathrm{d}r,
\quad
e^{(2)}=\sqrt{\Sigma}\:\!\mathrm{d}\theta,
\quad
e^{(3)}=\sqrt{\frac{A}{\Sigma}} 
\sin \theta \left(\mathrm{d}\varphi-\frac{2ar}{A}\:\!\mathrm{d}t\right).
\end{align}
Recall that the LNRF reduces to the static frame in the Schwarzschild case ($a=0$).
Figure~\ref{fig:LNRF} shows the comparisons 
between $P$ in Eq.~\eqref{eq:P} (solid curves) and 
the escape probability of a photon emitted from a rest source at the LNRF, 
$P_{\mathrm{L}}$ (dotted curves).
We can regard $P_{\mathrm{L}}$ as the photon escape probability 
that is not affected by the proper motion of an emitter. 
With $P_{\mathrm{L}}$ as the reference, for all $a$, 
we find that $P>P_{\mathrm{L}}$ in the early phase. 
We can interpret it as an increase in $P$ 
due to the effect of the proper $\varphi$-motion, 
i.e., the relativistic beaming toward spatial infinity occurs.
On the other hand, we find that $P<P_{\mathrm{L}}$ in the late phase. 
We can interpret it as the suppression of $P$ due to the increase of the velocity ratio 
\begin{align}
\nu=\big|
v_*^{(1)}/v_*^{(3)}
\big|
\end{align}
in the late phase,
where 
$v_*^{(i)}$ are the 3-velocity components relative to the LNRF, 
\begin{align}
v_*^{(i)}=\left.
\frac{u^a e^{(i)}_{a}}{u^a e^{(0)}_{a}}\right|_{r=r_*, \theta=\theta_*}.
\end{align}
The ratio $\nu$ monotonically increases 
from nearly zero at the initial point to infinity at the horizon. 
In other words, as the effect of the proper $r$-motion increases, 
the rate of the relativistic beaming toward the black hole increases, 
and as a result, $P$ is suppressed.
Let $r_{0, \mathrm{L}}$ be the radius at which $P_{\mathrm{L}}=1/2$. 
For $a\lesssim 0.401$, we have $r_0>r_{0, \mathrm{L}}$. 
This means that the proper motion of the emitter acts to suppress $P$ 
and makes a negative contribution to observability.
However, for $a\gtrsim 0.401$, we have $r_0<r_{0, \mathrm{L}}$. 
In other words, the proper motion of the emitter acts to increase $P$ 
and expands the observable region toward the horizon.

\begin{figure}[t]
\centering
 \includegraphics[width=7.8cm,clip]{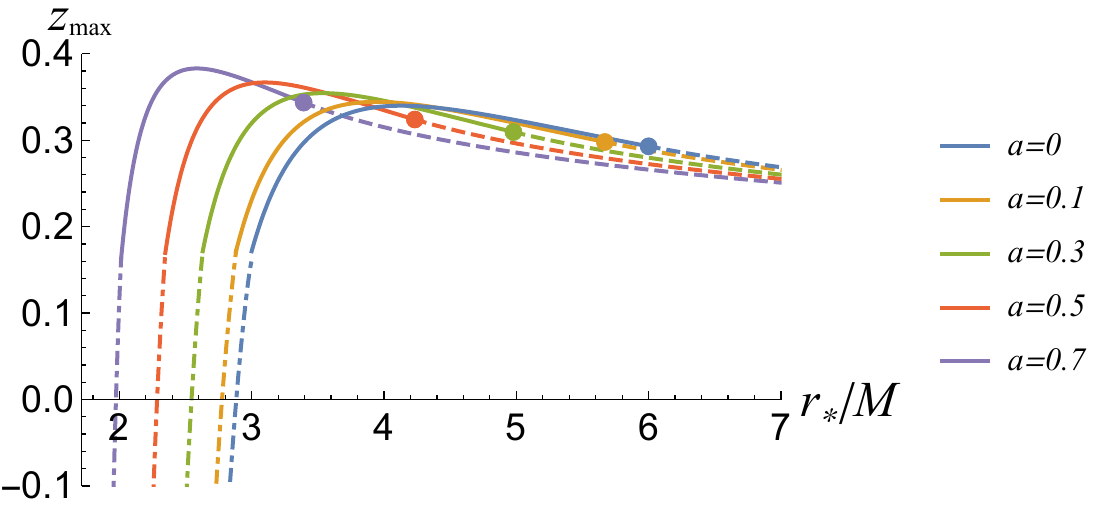}
 \includegraphics[width=8.4cm,clip]{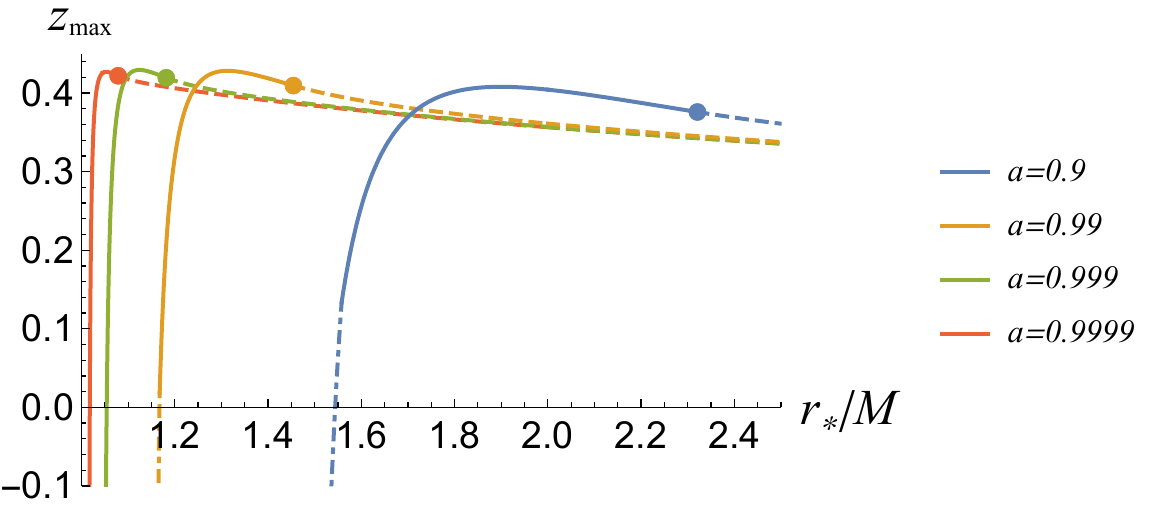}
 \caption{
 Maximum blueshift. 
 Solid curves: $z_{\mathrm{max}}$ 
 in the range $r_1^{\mathrm{c}}\leq r_*<r_{\mathrm{I}}$. 
 Dash-dotted curves: $z_{\mathrm{max}}$ 
 in the range $r_{\mathrm{H}}<r_*<r_1^{\mathrm{c}}$. 
 Dashed curves and dots: the maximum blueshift of an escape photon emitted 
 from an emitter orbiting circular geodesics and the ISCO, respectively. 
 }
 \label{fig:z}
\end{figure}

Let us focus on the frequency shift of escape photons. 
Let $z$ be the blueshift factor distribution of photons emitted from the emitter, 
\begin{align}
z=1+k_*^{(0)}.
\end{align}
Figure~\ref{fig:z} shows the maximum value $z_{\mathrm{max}}$ of the blueshift factor $z$, 
where the corresponding parameters are $(b, q, \sigma_r)=(b_1(r_*; 0), 0, 1)$ 
for $r_*\geq r_1^{\mathrm{c}}$ (solid curves) 
and $(b, q, \sigma_r)=(b_1^{\mathrm{c}}, 0, 1)$ for $r_{\mathrm{H}}<r_*<r_1^{\mathrm{c}}$ (dash-dotted curves). 
Dashed curves and dots in the range $r_*\geq r_{\mathrm{I}}$ also 
denote the maximum blueshift factor of an escape photon emitted from 
an emitter orbiting circular geodesics and the ISCO, respectively, 
where the corresponding parameters are $(b, q, \sigma_r)=(b_1(r_*; 0), 0, 1)$. 
Let $r_{\mathrm{z}}$ be the radius at which $z_{\mathrm{max}}=0$. 
For all $a$, we have $z_{\mathrm{max}}>0$ in the range $r>r_{\mathrm{z}}$, i.e., 
the energy of escape photons can blueshift there. 
On the other hand, we have $z_{\mathrm{max}}<0$ in the range $r<r_{\mathrm{z}}$. 
This means that all the escape photons are no longer blueshifted in this region.
Several values of $r_{\mathrm{z}}$ are shown in Table~\ref{table:rs} 
and are plotted in Fig.~\ref{fig:half}. 
We see that $r_0>r_{\mathrm{z}}$ for all $a$. 
This implies that the Doppler blueshift due to the emitter proper motion 
is dominant and can sufficiently cancel the gravitational redshift in the observation of the 
emitter in the range $r>r_0$. 

\section{Summary and discussions}
\label{sec:6}
We have considered isotropic photon emission from an emitter moving near the Kerr black hole. 
Clarifying necessary and sufficient escape conditions for photons emitted 
from the equatorial plane, we have specified whole parameter regions of photon escape 
in the two-dimensional impact parameter space
without restricting the radial position of the emission. 
We have focused on the dynamics of an emitter, which 
falls gently from the ISCO and plunges into the black hole horizon. 
Assuming initial conditions of the emitter to be 
adiabatically shifted slightly inward from the ISCO radius, 
we have obtained the unique plunge orbits analytically. 
In the early phase, the emitter orbits the black hole many times in the vicinity of the ISCO, 
and eventually, 
it turns to plunge toward the horizon in the late phase.

We have defined the escape probability $P$ of a photon isotropically emitted from the emitter
and have shown that $P$ decreases monotonically 
with the radial position of the emitter.
In this paper, we have supposed that the emitter is observable if $P>1/2$. 
For any black hole spin parameter, $P$ is greater than $1/2$ initially. 
As it falls toward the center and passes through $r=r_0$, 
the probability eventually becomes less than a half. 
Therefore, we have concluded that the emitter is observable 
until it reaches $r=r_0$, which is estimated as 
approximately intermediate value between the radii of the ISCO and the horizon. 
We have further clarified that $P$ depends on the proper motion of the emitter.
The probability is larger than the reference value $P_{\mathrm{L}}$ 
due to proper $\varphi$-motion in the early phase, while in the late phase, 
$P$ is smaller than $P_{\mathrm{L}}$ due to proper $r$-motion.
However, for $a\lesssim 0.401$, the proper motion acts to make $r_0$ larger; 
in other words, it makes a negative contribution to observability. 
On the other hand, for $a\gtrsim 0.401$, the proper motion acts to make $r_0$ smaller, 
which means that the proper motion positively affects the observability, 
and we can continue to observe the emitter closer to the horizon.
Furthermore, we have seen that the Doppler shift due to the proper motion can cancel 
the gravitational redshift and can cause blueshift.

The distribution of the blueshift factor may contain additional characteristics related to the proper motion, 
which will be reported in a separate paper in detail.%
\footnote{It was recently discussed that the maximum blueshift of a light source 
falling vertically in the radial direction in the Schwarzschild black hole spacetime is 
related to the light ring~\cite{Cardoso:2021sip}.} 
The quantities we have discussed here, the escape cone $\mathcal{E}$, 
the photon escape probability $P$, and the blueshift factor distribution $z$, 
provide a basis for evaluating observables such as energy fluxes. 
The evaluation and characterization of such physical quantities are left for future work.

We have assumed a small value for the radial initial velocity of our emitter, 
which is justified in the standard disk model. 
In the radiatively inefficient accretion flow model, 
we may have to assume a larger value for its initial velocity. 
This is also left for future work.

\begin{acknowledgments}
The authors are grateful to Takahiko Matsubara, Toyokazu Sekiguchi, 
Takahiro Tanaka, Norichika Sago, Keisuke Nakashi, Chul-Moon Yoo, Kouji Nakamura, Takaaki Ishii, Hiroyuki Nakano, and Hirotaka Yoshino for useful comments. 
This work was supported by JSPS KAKENHI Grants No.~JP19K14715 (T.I.), 
No.~JP20K14467, No.~JP20J00416 (K.O.), 
No.~JP17H01131, No.~JP19H05114, and No.~JP20H04750 (K.K.). 
\end{acknowledgments}

\appendix

\section{SCHWARZSCHILD CASE ($a=0$)}
\label{sec:A}

\subsection{PHOTON ESCAPE CONDITIONS}
\label{sec:A-1}
In the Schwarzschild case $a=0$, 
the radial equation of motion~\eqref{eq:rdot} reduces to 
\begin{align}
&\dot{r}^2+V(r)=\frac{1}{q+b^2},
\\
&V(r)=\frac{1}{r^2}-\frac{2}{r^3},
\end{align}
where we have rescaled $\tau$ by $(q+b^2)^{1/2}$. 
The function $V$ is the familiar effective potential of photon motion 
and is related to $b_i$ in Eqs.~\eqref{eq:b1} and \eqref{eq:b2} as 
$V=(q+b_i^2)^{-1}$. 
It takes a local maximum value $V=1/27$ at $r=3$ (i.e., $r_i^{\mathrm{c}}
=3$).
As a result, the photon escapable parameter region in Table~\ref{table:escapecond}
reduces to Table~\ref{table:a=0}, where 
we have used $q+(b_i^{\mathrm{s}})^2=27$ and $q_{\mathrm{max}}=V^{-1}(r_*)$.

\begin{table}[t]
\centering
\caption{Photon escape parameter region in the Schwarzschild spacetime ($a=0$).}
\begin{tabular}{lll}
\hline\hline
Case &$\sigma_r=+$&$\sigma_r=-$
\\
\hline
$2<r_*<3$~~~&$q+b^2<27$&n/a
\\
$r_*\geq 3$&$q+b^2\leq q_{\mathrm{max}}$~~~&$27<q+b^2<q_{\mathrm{max}}$
\\
\hline\hline
\end{tabular}
\label{table:a=0}
\end{table}

\subsection{Dynamics of an emitter}
\label{sec:A-2}
We consider the plunge orbit of Sec.~\ref{sec:4} in the case of the Schwarzschild spacetime. 
The energy and angular momentum of the emitter reduces to 
$E=2\sqrt{2}/3$ and $L=2\sqrt{3}$, respectively. 
Then, Eqs.~\eqref{eq:tdotpl}--\eqref{eq:udot} reduce to 
\begin{align}
&\dot{t}=\frac{E \:\!r }{r-2},
\\
\label{eq:phidots}
&\dot{\varphi}=\frac{L}{r^2},
\\
\label{eq:rdots}
&\dot{r}
=-\frac{1}{3}\left(\frac{6}{r}-1\right)^{3/2},
\end{align}
where we have used $r_{\mathrm{H}}=2$ and $r_{\mathrm{I}}=6$. 
Integrating these equations, we obtain
\begin{align}
&\frac{r_{\mathrm{I}}}{r}
=1+\frac{L^{2}}{(\varphi-\varphi_0)^2},
\\
\label{eq:tSsol}
&t-t_0=-\frac{2\sqrt{2}\left(24u-1\right)}{u\sqrt{6u-1}}
-44\sqrt{2} \tan^{-1}(\sqrt{6u-1})
+4\tanh^{-1}(\sqrt{(6u-1)/2}),
\end{align}
which correspond to Eqs.~\eqref{eq:tr} and \eqref{eq:phir} for $a=0$, respectively.
These results coincide with those of Ref.~\cite{Hadar:2009ip}.

\section{INGOING KERR-SCHILD COORDINATES}
\label{sec:B}
The dynamics of particles near the horizon is not well described 
in the Boyer-Lindquist coordinates because $(\mathrm{d}t)_a$ and $(\mathrm{d}\varphi)_a$ 
are singular in the horizon limit 
(i.e., the norms $\| \mathrm{d}t\|, \|\mathrm{d}\varphi \| \to \infty$ as $r\to r_{\mathrm{H}}$). 
As seen in Eqs.~\eqref{eq:tdot} and \eqref{eq:phidot} or 
Eqs.~\eqref{eq:tdotpl} and \eqref{eq:phidotpl}, 
even though $u^a$ is regular on the horizon, 
the components $\dot{t}=u^a (\mathrm{d}t)_a$ and $\dot{\varphi}=u^a (\mathrm{d}\varphi)_a$ 
diverge in the horizon limit. 
Therefore, it is useful to adopt the ingoing Kerr-Schild coordinates 
$(v, r, \theta,\psi)$, which are regular even on the future horizon, 
rather than the Boyer-Lindquist coordinates 
to describe the dynamics of particles near the horizon.
The coordinate transformation between these coordinates 
(see, e.g., Ref.~\cite{Poisson:2009pwt}) are given by
\begin{align}
v&=t+\int \frac{r^2+a^2}{\Delta}\:\!\mathrm{d}r
=t+r+\frac{M r_{\mathrm{H}}}{\sqrt{M^2-a^2}}\ln \left|
\frac{r}{r_{\mathrm{H}}}-1
\right|-\frac{M r_-}{\sqrt{M^2-a^2}}\ln \left|\frac{r}{r_-}-1\right|,
\\
\psi&=\varphi+\int \frac{a}{\Delta} \:\!\mathrm{d}r
=\varphi+\frac{a}{2\sqrt{M^2-a^2}} \ln \left|\frac{r-r_{\mathrm{H}}}{r-r_-}\right|.
\end{align}
Then, the metric in the ingoing Kerr-Schild coordinates is given by
\begin{align}
\mathrm{d}s^2
=-\left(1-\frac{2Mr}{\Sigma}\right) \:\!\mathrm{d}v^2
+2\:\!\mathrm{d}v\:\!\mathrm{d}r
+\Sigma\:\!\mathrm{d}\theta^2
+\frac{A}{\Sigma}\sin^2\theta\:\!\mathrm{d}\psi^2
-2a \sin^2\theta\:\!\mathrm{d}r\:\!\mathrm{d}\psi
-\frac{4Ma r}{\Sigma}\sin^2\theta \:\!\mathrm{d}v \:\!\mathrm{d}\psi.
\end{align}

\section{EVALUATION OF THE PHOTON ESCAPE PROBABILITY}
\label{sec:C}
We demonstrate one of the practical ways to evaluate $P$ in Eq.~\eqref{eq:P}. 
Let $(\alpha_{\mathrm{crit}}, \beta_{\mathrm{crit}})$ be 
the values of $(\alpha, \beta)$ evaluated with $\sigma_\theta=-1$, 
$b=b_{\mathrm{SPO}}(\xi)$, and $q=q_{\mathrm{SPO}}(\xi)$, 
which are the so-called critical angles of photon escape~\cite{Ogasawara:2019mir}.
We define a function $f_{\sigma_r}(r_*; \xi)$ in terms of these angles by
\begin{align}
f_{\sigma_r}(r_*; \xi)=
\frac{1}{2\pi}\cos \beta_{\mathrm{crit}}
\frac{\mathrm{d}\alpha_{\mathrm{crit}}}{\mathrm{d}\xi} 
=\frac{1}{2\pi}\frac{\cos\beta_{\mathrm{crit}}}{\cos\alpha_{\mathrm{crit}}}
\frac{\mathrm{d}\sin \alpha_{\mathrm{crit}}}{\mathrm{d} \xi}.
\end{align}
Then, the photon escape probability $P$ is written as 
\begin{align}
P(r_*; a)=p+\Theta(-p),
\end{align}
where $\Theta$ is a step function (i.e., $\Theta(\eta)=1$ 
for $\eta\geq 0$ and $\Theta(\eta)=0$ for $\eta<0$), and 
\begin{empheq}[left={p(r_*; a)=\empheqlbrace}]{alignat=3}
& \displaystyle\int_{r_2^{\mathrm{c}}}^{r_1^{\mathrm{c}}} f_{1} \:\!\mathrm{d}\xi
&&(r_{\mathrm{H}}<r_*<r_1^{\mathrm{c}}),
\\
&
\displaystyle\int_{r_2^{\mathrm{c}}}^{r_*} f_1\:\!\mathrm{d}\xi+
\displaystyle\int_{r_*}^{r_1^{\mathrm{c}}} f_{-1}\:\! \mathrm{d}\xi
&~~&(r_1^{\mathrm{c}}\leq r_*<r_2^{\mathrm{c}}),
\\
& \displaystyle\int_{r_2^{\mathrm{c}}}^{r_1^{\mathrm{c}}} f_{-1} \:\!\mathrm{d}\xi
&&(r_*>r_2^{\mathrm{c}}).
\end{empheq}


\end{document}